\documentstyle[eqsecnum,epsfig,psfig,aps]{revtex}

\def\eps{\varepsilon}
\def\Dm{\widetilde{\cal D}_{\mu}}
\def\D{{\cal D}}
\def\bfx{{\bf x}}

\begin{document}
\draft

\title{Field theoretic renormalization group
        for a nonlinear diffusion equation}

\author{N.~V.~Antonov$^1$ and Juha~Honkonen$^2$}
\address{$^1$Department of Theoretical Physics, St.~Petersburg University,
Uljanovskaja 1, \\ St.~Petersburg, Petrodvorez, 198504 Russia \\
$^2$Theory~Division, Department~of~Physical Sciences,
FIN-00014 University~of~Helsinki, Finland}

\date{\today}

\maketitle

\begin{abstract}
The paper is an attempt to relate two vast areas of the applicability
of the renormalization group (RG): field theoretic models and partial
differential equations. It is shown that the Green function of a
nonlinear diffusion equation can be viewed as a correlation function
in a field-theoretic model with an ultralocal term, concentrated at
a spacetime point. This field theory is shown to be
multiplicatively renormalizable, so that the RG equations can be
derived in a standard fashion, and the RG functions (the $\beta$ function
and anomalous dimensions)
can be calculated within a controlled approximation.
A direct calculation carried out in the two-loop approximation for the
nonlinearity of the form $\phi^{\alpha}$, where $\alpha>1$ is not
necessarily integer, confirms the validity and self-consistency of
the approach. The explicit self-similar solution is obtained for
the infrared asymptotic region, with exactly known exponents; its
range of validity and relationship to previous treatments are
briefly discussed.
\end{abstract}


\section{Introduction} \label{sec:Intro}

The renormalization group (RG) has proved to be the most efficient tool
for studying self-similar scaling behavior. First appeared within the
context of quantum field theory \cite{Bogol},
it was then successfully applied to a variety of problems as disparate
as phase transitions, polymer dilutes, random walks, hydrodynamical
turbulence, growth processes, and so on; see, e.g., the monorgaphs
\cite{Zinn,book3}, the proceedings \cite{Dubna}, and
references therein.

The most powerful and well-developed formulation of the RG is the field
theoretic one; see \cite{Bogol,Zinn,book3}. It is this version of the
RG that is simplest and most convenient in practical calculations,
especially in higher orders. It is also important that it has a reliable
basis in the form of quantum-field renormalization theory, including
the renormalization of composite operators and operator product expansion.
For this reason, the first step in the RG analysis of a given problem
is to reformulate it as a field theoretic model. This means that
the quantities under study should be represented as functional averages
with the weight $\exp S(\phi)$, where $\phi$ is a classical random
field (or set of fields) and $S(\phi)$ is certain action functional.
For parabolic differential equations with an additive random source,
such a formulation is provided by the well-known Martin--Siggia--Rose
formalism, see \cite{MSR,MSRF}. In problems involving
fluctuation effects in chemical reactions the somewhat more
complicated approach of Doi \cite{Doi76}
(see also \cite{Zeldovich78,Grassberger80}) has also been widely used
\cite{Peliti86,Lee94,Cardy9698}.
No general recipe, however, seems to exist to cast a nonlinear problem
to a field-theoretic form.

Such a reformulation, however, is by no means superfluous: once the field
theoretic formulation has been found, it becomes possible to apply standard
tools (power counting of the 1-irreducible correlation functions etc) to
verify the renormalizability of the model, i.e., the applicability of the RG
technique, to derive corresponding RG equations, and to calculate its
coefficients (beta functions and anomalous dimensions) within controlled
approximations. An instructive example is provided by the model of the
so-called true self-avoiding random walks \cite{tsaw,tsaw1,tsaw2}. After its
field theoretic formulation had been found \cite{tsaw1}, it became clear that
the model in its original formulation was not renormalizable, and the direct
application of the RG to it would lead to completely erroneous results. The
renormalizable version of the model can be obtained by adding of infinitely
many terms to the original action; see \cite{tsaw2}.

It has long been known, however, that symmetries of the RG type also
appear in various physical problems described by ordinary or partial
differential equations and integro-differential equations, whose solutions
exhibit self-similar scaling behavior \cite{Shirkov1}. Since then, the
list of such problems has been essentially increased; see, e.g.,
\cite{Shirkov3,Oono1,Bric1,Bric3,Bric4,Teo,Heisen,Kovalev1}
and references therein. As a rule, the field theoretic formulation for
these models does not exist (or, at least, is not known), and derivation
of the corresponding RG equations is a nontrivial task. Quoting the
authors of \cite{Kovalev1}, ``the procedure of revealing RG
transformations $\dots$
in any partial case $\dots$ up to now is not a regular one. In practice,
it needs some imagination and atypical manipulation
`invented' for every particular case.'' In Ref. \cite{Kovalev1}, a general
approach was proposed to constructing RG symmetries for certain
classes of partial differential equations, but its relationship to
the field theoretic RG techniques is not clear.

The present paper is an attempt to `bridge the gap' between these two
vast areas of the applicability of the RG: field theoretic models and
partial differential equations. To be specific, we shall
consider nonlinear diffusion equation of the form
\begin{equation}
\partial_{t} \phi = \nu_0 \partial^{2} \phi + V(\phi),
\label{diff}
\end{equation}
where $\phi(x)\equiv\phi(t,\bfx)$ is a scalar field, $\nu_0$ is the
diffusion coefficient, $\partial^{2}$ is the Laplace operator, and
$V(\phi)$ is some nonlinearity dependent on the field $\phi$ and its
spatial derivatives. Within the RG context, various special examples
of Eq. (\ref{diff}) were studied earlier in
\cite{Oono1,Bric1,Bric3,Bric4,Teo}.
In practical calculations, we shall confine
ourselves to the nonlinearity of the form $V(\phi)=-\lambda_{0}
\phi^{\alpha}$, where $\alpha>1$ is not necessarily integer.

We shall show that the problem (\ref{diff}) can be cast into
a field theoretic model and apply the standard RG formalism to it to
establish the scaling behavior and to calculate corresponding
anomalous dimensions. Then we shall discuss the range of applicability
of the results obtained and their relationship to the previous RG
treatments of the model.

 \section{Field theoretic formulation and renormalization
 of the problem} \label{sec:FT}

We begin the analysis of the Cauchy problem (\ref{diff}) with a
localized initial condition which corresponds to the equation
\begin{equation}
\partial_{t} G = \nu_0 \partial^{2} G + V(G) + \delta(x-x_{0})
\label{Green}
\end{equation}
for the Green function $G(x|e_{0})$. It will be shown later that the large-scale asymptotic behavior
of this problem survives for all integrable initial conditions (i.e. such that
$\int\!d{\bf x}\,\phi(0,{\bf x})$ converges).
In Eq. (\ref{Green}) we denote $\delta(x-x_{0})\equiv\delta(t-t_{0})\delta^{(d)}(\bfx-\bfx_{0})$, where
$d$ is the dimensionality of the $\bfx$ space, and $e_{0} = \{x_{0},
\nu_0, \lambda_{0} \}$ is the full set of parameters.

The functional derivation of the MSR formalism \cite{MSRF} can be
adopted to represent the solution of Eq. (\ref{Green}) as a functional
integral over the doubled set of fields, $\phi$ and $\phi'$:
\begin{equation}
G(x|e_{0}) = \int \D\phi' \int \D\phi\ \phi(x) \exp
\bigl[S(\phi',\phi) + \phi'(x_{0}) \bigr].
\label{funi}
\end{equation}
Here the normalization constant is included into the differential
$\D\phi' \D\phi$, the action functional has the form
\begin{equation}
S(\phi',\phi) = \int dx\ \phi'(x) \Bigl\{ -\partial_{t} \phi(x)+
 \nu_0 \partial^{2} \phi (x) + V\bigl(\phi(x)\bigr) \Bigr\},
\label{act}
\end{equation}
with $dx = dt\, d\bfx$. The last term in (\ref{Green}) can be treated
as an addition to the `interaction' $V(\phi)$ and gives rise to the
last term in the exponential of Eq. (\ref{funi}). The term quadratic
in $\phi'$, typical to the MSR actions, is absent in (\ref{act}) owing
to the absence of the random force in Eq. (\ref{diff}).

Representation (\ref{funi}) shows that the Green function (\ref{Green})
can be viewed as the correlation function $\langle \phi (x) \exp
\phi'(x_{0}) \rangle$ in the field-theoretic model with the action
(\ref{act}). It is not convenient, however, to deal with the
exponential composite operator $\exp \phi'$. A more useful
interpretation is the following: the integral (\ref{funi})
describes the correlation function $\langle \phi (x) \rangle$ for the
extended action $S'=S+\phi'(x_{0})$ with an `ultralocal' interaction
term concentrated on a single spacetime point $x_{0}$.

The renormalization of field theoretic models with ultralocal terms,
concentrated on surfaces, was studied in Ref. \cite{Symanzik} in
detail. The analysis of \cite{Symanzik}, which we also naturally
take to apply to our case, has shown that the standard renormalization
theory is applicable to such models, with some obvious modification
(see below).

The analysis of ultraviolet (UV) divergences is based on the analysis
of canonical dimensions; see \cite{Bogol,Zinn,book3}.  Dynamical
models of the type (\ref{act}), in
contrast to static models, have two scales, the length scale $L$ and
the time scale $T$. Therefore, the canonical dimension of
any quantity $F$ (a field or a parameter in the action functional)
is described by two numbers, the momentum dimension $d_{F}^{k}$ and
the frequency dimension $d_{F}^{\omega}$, determined so that
$[F] \sim [L]^{-d_{F}^{k}} [T]^{-d_{F}^{\omega}}$.
The dimensions are found from the obvious
normalization conditions $d_k^k=-d_{\bf x}^k=1$, $d_k^{\omega }
=d_{\bf x}^{\omega }=0$, $d_{\omega }^k=d_t^k=0$,
$d_{\omega }^{\omega }=-d_t^{\omega }=1$, and from the requirement
that each term of the action functional be dimensionless (with
respect to the momentum and frequency dimensions separately).
Then, based on $d_{F}^{k}$ and $d_{F}^{\omega}$,
one can introduce the total canonical dimension
$d_{F}=d_{F}^{k}+2d_{F}^{\omega}$ (in the free theory,
$\partial_{t}\propto\partial^{2}$), which plays in the theory of
renormalization of dynamical models the same role as the
conventional (momentum) dimension does in static problems;
see \cite{Zinn,book3}.

Now let us turn to the special case of the model (\ref{Green}) with
the extended action of the form
\begin{equation}
S'(\phi',\phi) = \int dx\ \phi'(x) \Bigl\{ -\partial_{t} \phi(x)+
 \nu_0 \partial^{2} \phi (x)  - g_{0}\nu_0\phi^{\alpha}(x) \Bigr\}
+ \phi'(x_{0}),
\label{extend}
\end{equation}
where we have introduced the new parameter $g_{0}\equiv\lambda_{0}/\nu_0$,
which plays the part of the coupling constant (a formal small parameter
of the ordinary perturbation theory).
Canonical dimensions for the model (\ref{extend}) are given in Table
\ref{table1}, including the dimensions of renormalized parameters,
which will appear later on. From Table \ref{table1} it follows
that the model is logarithmic (the coupling constant
$g_{0}$ is dimensionless) for $2+d(1-\alpha)=0$. In what follows,
we fix the exponent $\alpha$ in Eq. (\ref{extend}) and consider the
model in variable space dimension $d= (2-\eps)/(\alpha-1)$. Then the UV
divergences take on the form of the poles in
$\eps\equiv 2+d(1-\alpha)$ in the correlation functions.
The `interaction' is therefore irrelevant (in the sense of Wilson)
for $\eps<0$, marginal (logarithmic) for $\eps=0$, and relevant
for $\eps>0$; cf. the analysis in Ref. \cite{Bric1}. This means
that for $\eps\ge0$, the ordinary perturbation expansion (i.e.,
series in $g_{0}$) fails to give correct infrared (IR)
behavior and has to be summed up. The desired summation can be
accomplished using the renormalization group.

It is common wisdom of the renormalization theory that for the analysis of
UV divergences of all correlation functions of the
fields $\phi$ and $\phi'$ it is sufficient to consider one-particle-irreducible (1PI) correlation
functions, whose graphical representation contains only graphs which remain connected
after removal of one (arbitrary) line (i.e. a free-field correlation or response function) of the graph.

The total canonical dimension of an arbitrary 1PI correlation
function
\begin{equation}
\Gamma (x_{1},\cdots,x_{N};y_{1},\cdots, y_{N'};x_{0})
= {\delta^{N+N'}\Gamma(\phi,\phi')\over \delta\phi(x_{1}) \cdots\delta \phi (x_{N}) \,
\delta\phi'(y_{1}) \cdots \delta\phi' (y_{N'})}\,,
\label{GammaG}
\end{equation}
in the time--coordinate representation is given by the relation
\begin{equation}
d_{\Gamma} = N(d+2-d_{\phi})+N'(d+2-d_{\phi'}),
\label{deltac}
\end{equation}
where $N$ and $N'$ are the numbers of corresponding fields.
In (\ref{GammaG}) $\Gamma(\phi,\phi')$ is the (dimensionless)
generating functional of 1PI Green functions. It should be noted,
however, that due to the presence of the ultralocal term in the action,
the functional
$\Gamma(\phi,\phi')$ is {\em not} the Legendre transform of the functional
$W(J,J')=\ln {\cal G}(J,J')$, where
${\cal G}(J,J')=
\int \D\phi' \int \D\phi\ \exp
\bigl[S'(\phi',\phi) + J\phi+J'\phi') \bigr]$
is the generating functional of Green functions of the model. Moreover,
contrary to the usual field theories, the functional $\ln{\cal G}(J,J')$
does not include all connected graphs of ${\cal G}(J,J')$.
By definition of the generating functional,
the 1PI Green function with $N$ external $\phi$ legs and $N'$ external $\phi'$ legs may be obtained
by $N$ functional differentiations of $\Gamma(\phi,\phi')$
with respect to the field $\phi$ and $N'$ differentiations with respect to
$\phi'$. The canonical dimensions of the functional derivatives are
related to the dimensions of the corresponding fields as
$d^{k} [\delta/\delta\phi] = d - d^{k}_{\phi}$,
$d^{\omega} [\delta/\delta\phi] = 1 - d^{\omega}_{\phi}$,
and similarly for the auxiliary field $\phi'$. Then the total canonical dimension of
the function (\ref{GammaG}) in the frequency--momentum representation
(obtained by the Fourier transformation with respect to all
$N+N'$ independent differences of the time and coordinate arguments)
is obtained from (\ref{deltac}) by subtracting the term
$(N+N')(d+2)$ and has the form
\begin{equation}
d_{\Gamma}= - d_{\phi} N - d_{\phi'} N' = -dN,
\label{deltacC}
\end{equation}
where the data from Table \ref{table1} are used in the last equality.

The quantity (\ref{deltacC}) is the formal index of the UV divergence
for the function $\Gamma$. Like for usual (local) models,
superficial UV divergences, whose removal requires counterterms,
can be present only in those functions $\Gamma$ for which
$\delta \equiv d_{\Gamma} |_{\eps=0}$
is a non-negative integer; see \cite{Bogol,Zinn,book3}.

From Eq. (\ref{deltacC}) we conclude that for any positive $d$, such
divergences can exist only in the 1PI functions with
$N=0$ and arbitrary value of $N'$. For all these functions $\delta=0$,
that is, the divergences are logarithmic and the corresponding counterterms
in the frequency--momentum representation are constants.

At first glance, we have established that the model (\ref{extend})
requires infinitely many counterterms, and hence it is not renormalizable.
However, it turns out to
be sufficient to renormalize the 1PI Green function $\Gamma(x;x_0)$
only to render the model finite, as we shall now show.

The first few Feynman diagrams of $G$ are shown
in Fig.~I for $\alpha=2$; the symmetry coefficients are shown for general
$\alpha$ (it would be embarrassing to depict
the diagrams for fractional $\alpha$, but the idea is the same).
The lines with a slash denote the bare propagator
\begin{equation}
\label{bare}
\Delta(t,{\bf r})=\langle\phi\phi'\rangle_{0}= { \exp [
-r^{2}/4\nu_0 t]\over (4\pi\nu_{0}t)^{d/2}}\,,
\end{equation}
the end with a slash corresponds to the field $\phi'$, and the
end without a slash corresponds to $\phi$.
The initial (left) point in each diagram
corresponds to $x$, and the final (right) point with variable number
of attached lines corresponds to $x_{0}$. The crucial point is that,
as is easily seen from Fig.~I, all possible 1PI subdiagrams
entering into the diagrams of $G$ belong to the only 1PI
function $\Gamma(x;x_0)$; no other 1PI
functions are involved. The function $G$ appears to be
 `closed with respect to the renormalization,' i.e., we can
eliminate their UV divergences by the only counterterm corresponding
to its 1PI part $\Gamma(x;x_0)$.
\begin{figure}[t]
\begin{center}
\epsfig{file=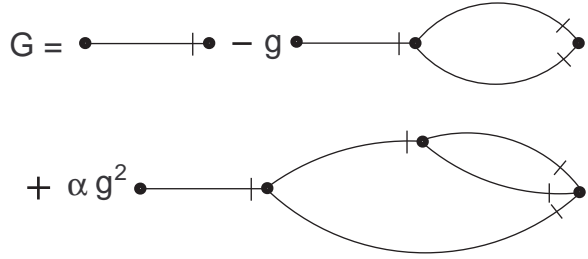,width=8 cm}\\
\end{center}
\caption{First Feynman graphs for the Green function of the nonlinear
diffusion equation (\ref{Green}) for $V(G)=-\lambda_{0} G^2$.}
\end{figure}
Moreover, the renormalization of the only function
$\Gamma(x;x_0)$ is in fact sufficient to completely
renormalize all functions with $N'>1$. A typical diagram for $N'=3$
is shown in Fig.~II. It is clear that any such diagram reduces to a
product of blocks that belong to the simplest function with $N'=1$
(we recall that there is no integration over $x_{0}$, the only point
that connects the blocks). Therefore, the diagram contains no superficial
divergences; all its divergences are those of the subdiagrams and they
are completely removed by the renormalization of the function with $N'=1$.
This is equally true for any diagram of any function with $N'>1$.

\begin{figure}[t]
\begin{center}
\epsfig{file=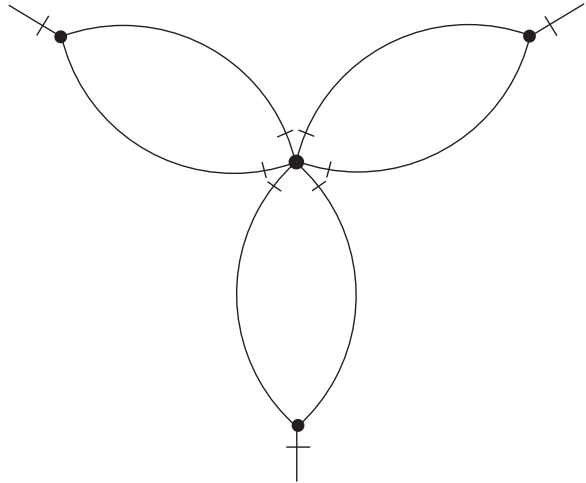,width=8 cm}\\
\end{center}
\caption{A three-loop Feynman graph for the three-point correlation function of the nonlinear
diffusion equation (\ref{Green}) for $V(G)=-\lambda_{0} G^2$ illustrating
the factorization property $
\Gamma(x_1,x_2,x_3;x_0)=\Gamma(x_1;x_0) \Gamma(x_2;x_0)\Gamma(x_3;x_0)$.}
\end{figure}

In the generic case all the loops are created by the presence of a single local
vertex with any number of $\phi'$ legs, from which continous chains of
retarded diffusion propagators emanate. Due to the structure of the nonlinear
term these chains do not branch, but they may merge (the single $\phi'$-field in the nonlinearity
allows only one outgoing propagator from each ordinary vertex, whereas up to
$\alpha$ incoming chains are allowed). A little reflection along the lines
sketched above shows then that
all divergent 1PI Green functions are factorized:
$$
\Gamma(x_1,\cdots,x_N;x_0)=\Gamma(x_1;x_0)\cdots \Gamma(x_N;x_0)\,.
$$
Thus, we are left with the only counterterm to the function
$\Gamma(x;x_0)$. It is constant (see above), which
in the time--coordinate representation corresponds to the function
$\delta(x-x_{0})\equiv\delta(t-t_{0})\delta^{(d)}(\bfx-\bfx_{0})$.
In the action functional, after the integration over the field argument,
this gives $\phi'(x_{0})$.
Such term is present in the extended action (\ref{extend}), so that
our model is renormalized multiplicatively,
with the only renormalization constant, which we denote $Z$. The
renormalized action has the form
\begin{equation}
S'_{R}(\phi',\phi) = \int dx\ \phi'(x) \Bigl\{ -\partial_{t} \phi(x)+
 \nu \partial^{2} \phi (x)  - g\nu\mu^{\eps}\phi^{\alpha}(x) \Bigr\}
+ Z\, \phi'(x_{0}).
\label{renac}
\end{equation}
Here and below the $g$ and $\nu$ are the renormalized analogs of the
bare parameters, $\mu$ is the reference mass in the minimal subtraction
(MS) scheme, which we use in practical calculations, and the
constant $Z$ depends on the dimensionless parameters $g$, $\alpha$
and $\eps$. The renormalized Green function $G_{R}$, which is finite
for $\eps\to0$, is given by the representation (\ref{funi}) with the
substitution $S'\to S'_{R}$.

If we now replace the local initial condition with an integrable one: $\phi(0,{\bf x})=a({\bf x})$,
then -- after Fourier transforming -- we obtain wave-vector integrals in which
all the propagator lines starting from the initial condition contain a multiplicative
factor $\tilde{a}({\bf k})=\int\!d{\bf x}\,e^{-i{\bf k}\cdot{\bf k}}a({\bf x})$. For the
large-scale asymptotic analysis using RG it is sufficient to keep the leading small wave-number
terms in all the lines which amounts to the replacement $\tilde{a}({\bf k})\to \tilde{a}(0)=
\int\!d{\bf x}\,a({\bf x})$, and we thus return to loop integrals of the
problem with localized initial condition in which $\int\!d{\bf x}\,a({\bf x})$ is
the amplitude of the initial $\delta$ function.

To clarify the idea, consider
the one-loop graph of Fig. I, whose analytic expression with the
initial condition $\phi(0,{\bf x})=a({\bf x})$ is
\[
\Gamma^{(1)}(t,{\bf x})=-\lambda_0\int\!d{\bf y}\!\int\limits_0^\infty\!dt'\Delta(t-t',{\bf x}-{\bf y})
\int\!d{\bf y}_1\Delta(t',{\bf y}-{\bf y}_1)a({\bf y}_1)\!
\int\!d{\bf y}_2\Delta(t',{\bf y}-{\bf y}_2)a({\bf y}_2)\,.
\]
Here, $\Delta(t, {\bf x})$ is the diffusion propagator (\ref{bare}).
Fourier transforming $\Gamma^{(1)}(t,{\bf x})$ with respect to ${\bf x}$ we arrive at the expression
\begin{equation}
\label{G01k}
\Gamma^{(1)}(t,{\bf k})=-\lambda_0
\int\limits_0^\infty\!dt'\Delta(t-t',{\bf k})\int\!{d{\bf q}\over (2\pi)^d}\Delta(t',{\bf q})\tilde{a}({\bf q})
\Delta(t',{\bf k}-{\bf q})\tilde{a}({\bf k}-{\bf q})\,,
\end{equation}
where $\Delta(t,{\bf k})$ is the spatial Fourier transform of the diffusion kernel (\ref{bare}).
From the point of view of RG, the IR relevant terms are given by the leading terms of
the gradient expansion of the initial condition: $\tilde{a}({\bf q})=\tilde{a}(0)+o(q)/q$.
This allows to replace (\ref{G01k}) by
\[
\Gamma^{(1)}(t,{\bf k})\sim-\lambda_0
\int\limits_0^\infty\!dt'\Delta(t-t',{\bf k})\int\!{d{\bf q}\over (2\pi)^d}\Delta(t',{\bf q})
\Delta(t',{\bf k}-{\bf q})\tilde{a}^2(0)\,,
\]
which corresponds to the localized initial condition with the amplitude
$\tilde{a}(0)=\int\!d{\bf x}\,a({\bf x})$.

 \section{RG equations and RG functions} \label{sec:RGE}

It follows from Eqs. (\ref{act}), (\ref{extend}), and (\ref{renac})
that the original and renormalized action functionals
satisfy the relation $S'(Z\phi', Z^{-1}\phi, e_{0})=
S'_R(\phi',\phi,e,\mu)$,
if the bare and renormalized parameters are related as follows:
\begin{equation}
 \nu_0 = \nu, \quad g_{0}=g \mu^{\eps} Z^{\alpha-1},
\label{reno}
\end{equation}
with the only renormalization constant $Z$ from Eq. (\ref{renac}).
This implies the relation $G(e_{0}) = Z^{-1} G_{R}(e,\mu)$
for the corresponding Green functions in Eq. (\ref{funi}); i.e.,
this quantity is multiplicatively renormalizable. We use
$\widetilde{\cal D}_{\mu}$ to denote the differential
operation $\mu\partial_{\mu}$ for fixed
$e_{0}$ and operate on both sides of this equation with it. This
gives the basic RG equation:
\begin{equation}
\Bigl[{\cal D}_{\mu} + \beta(g)\partial_{g} - \gamma(g)\Bigr]
G_{R}(e,\mu) = 0,
\label{RGE}
\end{equation}
where ${\cal D}_{\mu} + \beta(g)\partial_{g} $ is nothing else than
the operation $\widetilde{\cal D}_{\mu}$ expressed in the renormalized
variables.
In Eq. (\ref{RGE}), we have written ${\cal D}_{x}\equiv x\partial_{x}$
for any variable $x$, and the RG functions (the $\beta$ function and
the anomalous dimensions $\gamma$) are defined as
\begin{equation}
\gamma(g)\equiv \Dm \ln Z , \quad
\beta_{g}\equiv\Dm  g=g\Bigl[-\eps- (\alpha-1)\,\gamma(g) \Bigr].
\label{RGF}
\end{equation}
The relation between $\beta$ and $\gamma$ results from the
definitions and the relations (\ref{reno}).

We shall see below that, for small $\eps>0$, an IR stable fixed
point $g_{*}$ of the RG equation (\ref{RGE}) exists in the physical
region $g>0$ i.e., $\beta (g_*) = 0$, $\beta '(g_*) > 0$.
The functions $G$ and $G_{R}$ coincide up to a constant (i.e.,
independent of the time and space variables) factor $Z$ and the choice
of the parameters (bare $e_{0}$ or renormalized $e$, $\mu$) and can
equally be used in the analysis of the IR behavior.
The general solution of the RG equations is discussed in detail,
e.g., in \cite{Zinn,book3}. It follows from this solution that,
when an IR stable fixed point is present, the leading term of the
IR behavior of the function $G_{R}\propto G$ satisfies Eq. (\ref{RGE})
with the substitution $g\to g_{*}$:
\begin{equation}
\Bigl[{\cal D}_{\mu} - \gamma^{*}\Bigr] G_{R}(e,\mu) = 0.
\label{RGE2}
\end{equation}
In our case, the value of the anomalous dimension at the fixed point
is found exactly owing to the relation between $\beta$ and $\gamma$
in Eq. (\ref{RGF}):
\begin{equation}
\gamma^{*}\equiv\gamma(g_{*})=-\eps/(\alpha-1)=d-2/(\alpha-1).
\label{exact}
\end{equation}
Dimensional considerations yield $G_{R}(t,r)=(\nu t)^{-d/2}\xi(1/t\mu^{2}\nu,r^{2}/t\nu)$,
where $\xi$ is some function of dimensionless variables. The dependence on
$g$ is not displayed explicitly, because the derivatives with
respect to this parameter do not enter into Eq. (\ref{RGE2}).
It follows from Eq. (\ref{RGE2}) that $\xi$ satisfies --- at the fixed point ---
the equation $\Bigl[{\cal D}_{s} - \gamma^{*}/2\Bigr] \xi(s,y)=0$,
its general solution is $\xi(s,y) = s^{\gamma^{*}/2} \chi(y)$,
where $\chi$ is an arbitrary function of the second variable $y$.
For the Green function (\ref{funi}) we then obtain
$$G(t,r)\sim G_{R}(t,r) \sim t^{-d/2+\gamma^{*}/2}
\chi(r^{2}/t\nu)= t^{-1/(\alpha-1)} \chi(r^{2}/t\nu), $$
where the form on the `scaling function' $\chi(r^{2}/t\nu)$ is not
determined by the equation (\ref{RGE2}). The dependence on the
parameters $\nu$, $\mu$
can be easily restored from the dimensionality considerations
(see Table I):
\begin{equation}
G(t,r)\sim ( \nu_0\,  t)^{-1/(\alpha-1)} \,
\chi(r^{2}/t\nu_0).
\label{solution}
\end{equation}
Although the value of $\gamma^{*}$ in Eq. (\ref{exact}) and the
solution (\ref{solution}) have been obtained
without practical calculation of the constant $Z$ and functions
(\ref{RGF}), such calculation is needed to check the existence,
positivity and IR stability of the fixed point. Within the $\eps$
expansion, these facts can be verified already in the simplest
one-loop calculation.

In order to check the validity and
self-consistency of the approach, we calculated the constant $Z$
up to the two-loop approximation. The calculation is performed in the
frequency--momentum ($\omega$, $k$) representation and calls for the
formulas derived in Ref. \cite{AV} for a model of critical dynamics.

Two key points are as follows: the convolution of two functions of the form
$F(\alpha;a) \equiv (-{\rm i}\omega\, a + k^{2})^{-\alpha}$
is a function of the same form,
$$F(\alpha;a) * F(\beta;b) =  K(\alpha,\beta;a,b)\,
F(\alpha+\beta-d/2-1;a+b)  $$
where $a$  and $b$ are both positive and the coefficient has the form:
$$ K(\alpha,\beta;a,b) = a^{d/2-\alpha} b^{d/2-\beta}
(a+b)^{\alpha+\beta-d-1}
\Gamma(\alpha+\beta-d/2-1) / \Gamma(\alpha) \Gamma(\beta) (4\pi)^{d/2}, $$
while the product of two such functions can be represented as a single
integral of a function of the same form with the aid of the generalized
Feynman formula:
$$ F(\alpha;a) \cdot F(\beta;b) = \frac{\Gamma(\alpha+\beta)}
{\Gamma(\alpha)\Gamma(\beta)} \, \int_{0}^{1} ds\,
s^{\alpha-1} (1-s)^{\beta-1} \, F(\alpha+\beta; as+b(1-s)). $$
For the sake of brevity, below we give only the final result:
\begin{equation}
Z = 1+ \frac{u}{\eps} + \frac{\alpha u^{2}} {2\eps^{2}}
-\frac{u^{2}} {2\eps} \, \tilde I_{\alpha} + O(u^{3}),
\label{Z}
\end{equation}
where we have introduced a new coupling constant,
\begin{equation}
u \equiv \frac{g}{(4\pi)}\, \alpha^{-1/(\alpha-1)}
\label{u}
\end{equation}
and have written
$\tilde I_{\alpha}\equiv \alpha\ln \alpha + \alpha^{\alpha/(\alpha-1)}
I_{\alpha}$ with the convergent single integral
\begin{eqnarray}
I_{\alpha}\equiv \int_{0}^{1} \frac{ds}{s} \biggl\{ \Bigl(
s(\alpha-1)+1 \Bigr)^{(2-\alpha)/(\alpha-1)} \Bigl( (s+1)(\alpha-1)+1
\Bigr) ^{-1/(\alpha-1)}
-\alpha^{-1/(\alpha-1)} \biggr\},
\label{In}
\end{eqnarray}
in particular, $\tilde I_{2}=2 \ln(4/3)$ and
$\tilde I_{3}=6 \bigl[\ln(3-\sqrt 5)+\ln(3/2)\bigr]$.

Then for the corresponding beta function we obtain
$\beta_{u}\equiv \Dm u = -u \bigl[ \eps + \beta_{u} \partial_{u}
\ln Z^{\alpha-1}\bigr]$, where we have used the last relation in Eq.
(\ref{reno}) and the fact that $\Dm = \beta_{u} \partial_{u}$ for the
functions dependent only on $u$. This yields
\begin{eqnarray}
\beta_{u}(u)= \frac {-\eps\,u}{1+ (\alpha-1)\,\D_u \ln Z}.
\label{beta1}
\end{eqnarray}
Substituting Eq. (\ref{Z}) into Eq. (\ref{beta1}) gives
\begin{equation}
\beta_{u}(u) =-u \bigl[ \eps -u(\alpha-1)+u^{2}(\alpha-1)
\tilde I_{\alpha} \bigr] + O(u^{4}).
\label{beta}
\end{equation}
Note that the poles in $\eps$ in the constant $Z$ cancel out in
the function (\ref{beta}); this is a manifestation of the general
fact that the RG functions must be UV finite, i.e., finite as
$\eps\to0$. The cancellation is possible by virtue of the correlation
that exists between the $u/\eps$ and $(u/\eps)^{2}$ terms in Eq.
(\ref{Z}) and can be used as an additional check of the consistency
of the approach. The simple (linear) dependence on $\eps$ is a
feature specific to the MS scheme.

From Eq. (\ref{beta}) we find an explicit expression for the
coordinate of the fixed point:
\begin{equation}
u_{*} = \frac{\eps}{(\alpha-1)} +  \tilde I_{\alpha}\,
\frac{\eps^{2}}{(\alpha-1)^{2}} +O(\eps^{3}).
\label{fipo}
\end{equation}
As already said above, for small positive $\eps$ and $\alpha>1$ the fixed
point is positive and IR stable: $\beta_{u}'(u_{*}) =\eps +O(\eps^{2})$.

\section {Discussion}

We have applied the field theoretic renormalization group to
the non-stochastic differential equation (\ref{Green}) and established the
scaling behavior in the IR asymptotic range, as a consequence of the
existence of the IR stable fixed point in the physical range of parameters.
The same asymptotic behavior is shown to be valid for integrable initial
conditions which thus constitute the universality class of this fixed point.

The key points are the formulation of the problem as a field theoretic model
with an ultralocal term concentrated at a spacetime point and the fact that
this model appears multiplicatively renormalizable, in spite of the naive
power counting that indicates nonrenormalizability.

The two-loop calculation confirms internal consistency of the approach.

Simple explicit form of the scaling dimensions follows from the fact
that there is only one independent renormalization constant in the problem.
In particular, this explains a simple value $z=2$ of the exponent in
the argument $r^{2}/t^{1/z}$ of the scaling function (\ref{solution})
(in models of dynamical critical phenomena \cite{Zinn,book3}
and some models of nonlinear diffusion \cite{JETP} this exponent
differs from two).

Recently, it has been conjectured \cite{Teo}
that the dynamic exponent $z\ne 2$ in the present problem. Our asymptotic solution
(\ref{solution}), however, does not predict any deviation from the canonical
value $z=2$, since there is no renormalization of the diffusion coefficient in
the MS scheme we have used. In Ref. \cite{Teo} with the use of a different renormalization
procedure it was concluded that $z-2=O(\varepsilon^2)$. We think, however,
that it is not consistent to prescribe physical quantities values of the order $O(\varepsilon^2)$
on the basis of the {\em one-loop} calculation carried out in Ref. \cite{Teo}, but
a two-loop analysis is required for this accuracy.


The RG analysis allows one to derive the RG equation rigorously and to
prove that the behavior (\ref{solution}) is indeed realized for
$\eps>0$, $g_0>0$ in the IR asymptotic range, specified by the
relations $t \sim r^{2}$ and $r\ll \eta$, where
$\eta\simeq g_{0}^{-1/\eps}$ is the UV scale. The general solution
of Eq. (\ref{RGE}) interpolates between the ordinary perturbation
theory for Eq. (\ref{Green}) and the self-similar asymptotic
expression (\ref{solution}). The scaling function $\chi(y)$ can be
calculated within the $\eps$ expansion; in the lowest order one
easily obtains $\chi(y)= \exp [-(y/2)^{2}]+O(\eps)$.

We hope that the ideas presented above might be useful in other models
containing ultralocal contributions, which have several charges and hence
richer IR behavior. Another direction of generalization would be the
analysis of Green functions of vector quantities.

\acknowledgments
We thank L.~Ts. Adzhemyan, A.~Kupiainen, M.~Yu.~Nalimov and A. N. Vasil'ev
for discussions.
The work was supported by the Grant Center for Natural Sciences (Grant No.
E00-3-24), the Nordic Grant for Network Cooperation with the Baltic
Countries and Northwest Russia No.~FIN-18/2001, and
the Academy of Finland (Grant No.~79781).

\begin{table}
\caption{Canonical dimensions of the fields and parameters in the
model (\protect\ref{act}).}
\label{table1}
\begin{tabular}{cccccc}
$F$ & $\phi$ & $\phi'$, $g$ & $\nu$, $\nu _{0}$ & $\mu$ & $g_{0}$ \\
\tableline
$d_{F}^{k}$ & $d$ & 0 & $-2$  & 1 & $2+d(1-\alpha) \equiv \eps $ \\
$d_{F}^{\omega }$ & 0 & 0 & 1 & 0 & 0 \\
$d_{F}$ & $d$ & 0 & 0 & 1 & $\eps$ \\
\end{tabular}
\end{table}

\begin{references}

\bibitem{Bogol} N. N. Bogoliubov and D. V. Shirkov,
{\it Introduction to the Theory of Quantized Fields}
(Wiley, New York, 1980).

\bibitem{Zinn} J. Zinn-Justin, {\it Quantum Field Theory and
Critical Phenomena} (Clarendon, Oxford, 1989).

\bibitem{book3} A. N. Vasil'ev,  {\it Quantum-Field Renormalization
Group in the Theory of Critical Phenomena and Stochastic Dynamics}
(St.~Petersburg Institute of Nuclear Physics, St.~Petersburg, 1998)
[in Russian; English translation: Gordon \& Breach, in preparation].

\bibitem{Dubna}
Proceedings~of~the International Conference
``Renormalization Group.''
D.~V. Shirkov, D.~I. Kazakov, and A.~A. Vladimirov (eds.),
World Scientific, 1988;
Proceedings of the Second International Conference
``Renormalization Group' 91.''
D.~V. Shirkov and V.~B. Priezzhev (eds.),
World Scientific, 1991;
Proceedings of the Third International Conference
``Renormalization Group' 96.''
D.~V. Shirkov, D.~I. Kazakov, and V.~B. Priezzhev (eds.),
JINR, Dubna 1997.

\bibitem{MSR} P. C. Martin, E. D. Siggia, and H. A.
Rose, Phys. Rev. A {\bf 8}, 423 (1973).

\bibitem{MSRF} H. K. Janssen, Z. Phys.
B {\bf 23}, 377 (1976); R. Bausch, H. K. Janssen, and H. Wagner,
Z. Phys. B {\bf 24}, 113 (1976); C. De Dominicis, J. Phys. (Paris)
{\bf 37}, Colloq. {\bf C1}, C1-247 (1976).


\bibitem{Doi76}
M. Doi, J. Phys. A: Math. Gen. {\bf 9}, 1465 (1976); {\bf 9}, 1479 (1976).

\bibitem{Zeldovich78}
Ya.B. Zel'dovich and A.A. Ovchinnikov, Zh. {\'E}ksp. Teor. Fiz.
{\bf 74}, 1588 (1978).

\bibitem{Grassberger80}
P. Grassberger and M. Scheunert, Fortschr. Phys. {\bf 28}, 547 (1980).

\bibitem{Peliti86}
L. Peliti,
J. Phys. A: Math. Gen. {\bf 19}, L365 (1986).

\bibitem{Lee94}
B. P. Lee,
J. Phys. A: Math. Gen. {\bf 27}, 2633 (1994).

\bibitem{Cardy9698}
J. L. Cardy and U. C. T\"auber, Phys. Rev. Lett. {\bf 77}, 4780 (1996); J. Stat. Phys. {\bf 90}, 1 (1998).






\bibitem{tsaw} D. J. Amit, G. Parisi, and L. Peliti,
Phys. Rev. B {\bf 27}, 1635 (1983).

\bibitem{tsaw1} S. P. Obukhov and L. Peliti, J. Phys. A: Math. Gen.
{\bf 16}, L147 (1983); S. A. Bulgadaev and S. P. Obukhov, Phys. Lett.
{\bf 98A}, 399 (1983); L. Peliti, Phys. Rep. {\bf 103}, 225 (1984).

\bibitem{tsaw2} S. \'{E}. Derkachov, J. Honkonen, and A. N. Vasil'ev,
J. Phys. A: Math. Gen. {\bf 23}, 2479 (1990).

\bibitem{Shirkov1} D. V.~Shirkov, Sov. Phys. Dokl. {\bf 27}, 197 (1982);
Theor. Math. Phys. {\bf 60}, 778 (1984).

\bibitem{Shirkov3} D. V.~Shirkov, Int. J. Mod. Phys. A {\bf 3}, 1321 (1988).

\bibitem{Oono1} N.~Goldenfeld, O.~Martin, and Y.~Oono,
J. Sci. Comp. {\bf 4}, 355 (1989);
N.~Goldenfeld, O.~Martin, Y.~Oono, and F.~Lui,
Phys. Rev. Lett. {\bf 64}, 1361 (1990).

\bibitem{Bric1} J.~Bricmont and A.~Kupiainen, Comm. Math. Physics {\bf 150},
193 (1992); Institut Mittag-Lefler Report No. 5, 1994/95.

\bibitem{Bric3} J.~Bricmont, A.~Kupiainen, and G.~Lin,
         Comm. Pure Appl. Math. {\bf 47}, 893 (1994).

\bibitem{Bric4} J.~Bricmont, A.~Kupiainen, and J.~Xin,
         J. Diff. Eqs. {\bf 130},  9 (1996).

\bibitem{Teo}  \'{E}. V. Teodorovich, Zh. {\'E}ksp. Teor. Fiz.
{\bf 115}, 1497 (1999).

\bibitem{Heisen}  L. Ts. Adzhemyan and N. V. Antonov,
Theor. Math. Phys. {\bf 115}, 562 (1998).

\bibitem{Kovalev1} V. F.~Kovalev, V. V.~Pustovalov, and D. V.~Shirkov,
        J. Math. Phys. {\bf 39}, 1170 (1998).

\bibitem{Symanzik} K. Symanzik, Nucl. Phys. B {\bf 190}, 1 (1981).

\bibitem{AV} N. V. Antonov and A. N. Vasil'ev,
Theor. Math. Phys. {\bf 60}, 671 (1984).

\bibitem{JETP} A. N. Vasil'ev, M. M. Perekalin, and A. S. Stepanenko,
       Zh. {\'E}ksp. Teor. Fiz. {\bf 100}, 1781 (1991).





\end{references}
\end{document}